# Developing Knowledge States: Technology and the Enhancement of National Statistical Capacity

**Derrick M. Anderson**
**Assistant Professor**
**School of Public Affairs**
**Arizona State University**
derrick.anderson@asu.edu

**Andrew Whitford**
**Alexander M Crenshaw Professor of Public Policy**
**Department of Public Administration and Policy**
**School of Public and International Affairs**
aw@uga.edu

**November 2014**

## Abstract

National statistical systems are the enterprises tasked with collecting, validating and reporting societal attributes. These data serve many purposes – they allow governments to improve services, economic actors to traverse markets, and academics to assess social theories. National statistical systems vary in quality, especially in developing countries. This study examines determinants of national statistical capacity in developing countries, focusing on the impact of general purpose technologies (GPTs). Just as technological progress helps to explain differences in economic growth, states with markets with greater technological attainment (specifically, general purpose technologies) arguably have greater capacity for gathering and processing quality data. Analysis using panel methods shows a strong, statistically significant positive linear relationship between GPTs and national statistical capacity. There is no evidence to support a non-linear function in this relationship. Which is to say, there does not appear to be a marginal depreciating National Statistical Capacity benefit associated with increases in GPTs.

## Introduction

Effective systems of government—tasked broadly with guiding the interrelated processes of formulating and implementing laws, rules and policies—rely heavily on information, data and evidence. Accordingly, support for systems to collect, validate and report information and data about a nation's population, resources and economy has existed for generations. For example, Article I, Section II of the US Constitution requires a national census of the population and Title 13 of the US Code enables the Census Bureau to do so. Arguably, the demands for high quality national data have increased in recent decades (World Bank 2002). Most of these data are produced by national enterprises generally referred to as national statistical systems. The quality of these systems varies widely even across nations of similar wealth and governance structure (Jerven 2013). National statistical systems are likely to emerge as important components of administrative states given the increasingly complex and information driven functions of modern governments.

The poor quality of national statistical systems in developing countries has long been recognized (World Bank 2002) and the widespread implications of this are readily apparent. For example, in 2010, the national statistical system of one developing country, Ghana, erroneously estimated their GDP and issued a revision that raised the statistic by some 60% (Jerven 2013, Devarajan 2013), the effect of which being an overnight reclassification of Ghana from a low-income to middle-income country (Jerven 2013); the implications of these issues are profound for foreign aid, lending, commerce, development and foreign investment. The response from the international economic development community was a call for serious reflection on Africa's so-called "statistical tragedy" (Devarajan 2013) and reinforced support for international efforts to evaluate and improve national statistical systems (e.g., Willoughby 2008). Thus, serious theoretical treatment of the determinants of quality in national statistical systems has considerable practical value. As one scholar observed, national statistics can be observed as problems of both "governance and knowledge" (Jerven 2013, S19).

The ambition in this study is to add to what is known about the determinants of national statistical capacity. This study considers technological capacity as a predictor of national statistical capacity. Economists view variance in technological progress as a critical driver of cross-national differences in economic growth (e.g., Helpman 1998; Islam 2003; Castellacci 2007). Technology also contributes to the development of governance practices and the regulatory state (Whitford and Tucker 2009). Accordingly, this paper contends that countries with more widespread use of "general purpose technologies" (GPTs) have greater national statistical capacities. The effects of GPTs on organizations and economic development are well documented. GPTs are innovations that may be used broadly throughout an economy, affecting the operations of a wide variety of organizations (Helpman 1998, 3); organizations that adopt these "enabling technologies" can quickly modify their activity to capitalize on market complementarities (Bresnahan and Trajtenberg 1995).

Evolutionary theories of economic development view technologies as infused within organizations, facilitating information exchange and increasing flexibility of organization level processes, thereby increasing competitiveness (Nelson and Winter 1982; Andersen 1994; Porter 1980; Utterback 1994). Technological advancement is the foundation for widespread economic development (WCED 1987; Dubose, et al. 1995; Sen 1999; Solow 1956, 1957). This line of thinking about GPTs' benefits is extended to governance structures, specifically national statistical capacity. The focus on the relationship between national statistical systems and technology addresses an important public administration perspective on technology in governance that is now decades old. This perspective holds that government support for technology development is rooted in national power; manifest early in technology for defense and today in technology for economic prosperity (Lambright 1989). The focus on national statistical systems, which tend to be used to support national wealth generating enterprises, provides a helpful lens for viewing the overwhelmingly complex interactions of technology, governance and economic activity. This paper tests the hypothesis—that countries with greater levels of general purpose technological attainment have higher quality national statistical systems—using cross-national data measured across time.

The focus of the present study is on a main regime for assessing national statistical capacity: the World Bank's Statistical Capacity Index (SCI). SCI has three components: statistical methodology, source data, and periodicity and timeliness. Statistical methodology refers to a country's adherence to international methodological standards. Source data measures a country's frequency of obtaining statistical data such as population and poverty rate. Periodicity and timeliness measures the availability of socioeconomic indicators, including those associated with the Millennium Development Goals. This component also rates the ease of access to key statistics. Countries were scored on each component on a scale of zero to one hundred. SCI is the mean of all three scores. This study uses a model based on panel data from 94 countries from 2004 to 2006, using national statistical capacity (measured through SCI) as the dependent variable. Findings from the model indicate that countries with greater levels of general purpose technological attainment have greater national statistical capacity.

This paper continues a review of the literature on technological achievement and mechanisms like national statistical capacity relative to our stated hypothesis. Next is a brief discussion of how core theories are operationalized before transitioning to a discussion of the multivariate estimation strategy and a review of the empirical results.

## General Purpose Technological Attainment

Technology is a major source of social change. One perspective on the role of technology in political and cultural change holds that while technology increases capacity for control it also ushers in economic, political and cultural complexity and uncertainty (LaPorte 1971). This study's thinking about technology and governance comes, at least in part, from economic considerations of the effects of technologies on organizations and nations. Over the last half-century, economists have understood that knowledge of how countries expand technological capacity can help explain economic growth (Schumpeter 1934; Solow 1956, 1957; Fagerberg 1994; Sen 1999). Endongenous growth theory, for example, posits that better technology caused the improved living standards experienced after the Industrial Revolution (Grossman and Helpman 1994). In 1957, Solow showed that seven-eighths of the increase in productivity in the U.S. economy was caused by technical change, including educational improvements and hard technologies. In contrast, expansion of labor and capital, long thought to be the main drivers of growth, explained very little (Solow 1957; see also Maddison 1987). One view is that organizations "learn by doing," and that knowledge differences affect the landscape of labor and capital across countries (Arrow 1962; Romer 1986; Lucas 1988). A parallel view is that organizations make investments in technology (in addition to human capital and knowledge) based on expected monetary gains (Grossman and Helpman 1994). In some cases, an organization's investment may expand the societal knowledge base, resulting in a public return that exceeds the organization's private return. Public policy helps explain whether organizations or institutions decide to make such investments (Romer 1994).

Technology can affect a nation's economy in two ways: 1) a firm gains private benefit from their investment, and 2) the effect of that investment increases the overall knowledge base of society (also called "knowledge capital"). Some innovations are particularly likely to have effects across the economy. "General purpose technologies" (GPTs) influence a variety of industries and supplement other, more specialized innovations (Bresnahan and Trajtenberg 1995). GPTs can increase knowledge capital and help organizations innovate, but they may provide more benefit to some sectors than others, just as some countries may be more likely to adopt them (Helpman 1998, 2).

Empirical studies demonstrate the ability of technology to help firms compete at both the national and global scale (Fagerberg 1994). Since technology varies across geographic regions, the level of technological capability across firms and organizations will vary as well (Solow 1960). In markets with minimal penetration of GPTs, firms lack the incentives to invest in new technologies that spur growth (Stiglitz 1989, see also Pietrobelli 1994). Practically, the economic focus – both theoretical and empirical – on technological capacity has led people to research how technology differs across nations (e.g., Archibugi and Coco 2005). The primary focus of this research has been on measuring national variation in access to technologies, such as higher education or GPTs.

The present case takes the position that countries with more access to GPTs also have greater national statistical capacity. It is axiomatic that information is needed for effective governance and policy making. National statistical systems are the organizations within governments that manage the processes of collecting, verifying and distributing statistical information about the country.

Specifically, organizations that can adopt technologies (those with "absorptive capacity") are able to invest in GPTs, along with more specialized technologies in order to improve performance (Cohen and Levinthal 1989, 1990). Cohen and Levinthal define absorptive capacity as the "ability to assimilate and exploit existing information" (1989, 569). Rather than learning-by-doing, organizations can gather outside information and use it to modify their approach. Organizations with greater absorptive capacity are more able to adopt GPTs at lower costs than those with less, allowing them to more easily capture the societal benefits of the technology. In this light, absorptive capacity and the ability to capture the benefits of technological advancement are "complementary assets" (Teece 1986).

Some economic perspectives of technology and organizations hold that organizations may benefit from technology through lower costs of production and better quality products (Porter 1980; Utterback 1994). Organizations will alter patterns of technology adoption in the face of future uncertainty (Dosi 1988); but adopted technologies become embedded and help organizations expedite information exchange and adapt more quickly (Nelson and Winter 1982; Andersen 1994). The insights on organizational aspects of technology are not lost in theoretical and practical reflections on national statistical systems. One expert in the area of national statistical capacity enhancement (Mizrahi 2004) notes that a dominant mode for increasing national statistical capacity is to focus on human capital but that providing adequate resources including information technologies and equipment are critical.

Organizations' ability to invest in beneficial production technology is bounded by their institutional absorptive capacity and the availability of GPTs that make specific higher-level technologies and practices possible. In seeking a competitive advantage, firms can seek to implement changes to their production process. Even those changes that only affect their administrative system are fundamental technological changes. The hypothesize taken here is that, like firms, nations with more absorptive capacity and greater penetration of GPTs are better able to support national statistical systems seeking to gather, validate and disseminate high quality statistical information. The hypothesis is as follows: *Countries with higher levels of technological attainment have higher levels of national statistical capacity*.

## National Statistical Capacity

National, international and transnational data play an increasingly important role in policymaking and governance at national and international levels (World Bank 2002). Much of these data is generated by national statistical systems. For instance, GDP is perhaps the most recognizable measure of a nation's economic condition. There are three primary sources for GDP: the World Bank's World Development Indicators, University of Pennsylvania's Penn World Tables, and the Madison data set—all measures that aggregate national level data that is collected and reported by national statistical systems (Jerven 2013).

The structure, operations and influences of national statistical systems have been examined from a host of theoretical and practical perspectives. Thus, due to the accumulation of research, it is anticipated that statistical capacity is associated with a number of important governance concepts and functions including bureaucratic quality (Williams 2006), government transparency (Williams 2011) and the quality of tax collection systems (Martin et al., 2009). There is some evidence that a nation's statistical capacity influences its ability to monitor critical health issues (Liberman, 2007, 1407) and plays an important role in tracking and understanding important and complex environmental issues (Gössling et al. 2002, 199). Political scientists have argued that national statistical capacity is associated with voice and accountability in government, political stability, government effectiveness, regulatory quality, and corruption (Angel-Urdinola et al. 2011). While many of these concepts and functions relate centrally to democratic governance, the virtues of a robust national statistical system may transcend the boundaries from democratic to authoritarian regimes as there is a least some evidence that even authoritarian regimes support improvements in national statistical capacity in efforts to enhance the likelihood of regime survival through, among other channels, participation in international efforts (Boix and Svolik 2013).

At the international level, the statistics created by national statistical systems play a critical role in international poverty reduction programs (Deaton 2001). They are, according to one expert assessment, "the starting point" in the war on poverty (World Bank 2010). They also contribute critically to monitoring the implementation of assorted international treaties related to peace and security (Sanga et al. 2011, 304).

Studies show considerable variation in national statistical capacity even among countries of similar wealth and in the same geographic regions (Jerven 2013). Thus, the economic development community knows that, as one report observed, "the nature and organization of national statistical agencies vary according to the political system, the demand for data, and the organization of local and central governments" (World Bank 2002). There is a growing recognition of the need for improved national statistical capacity in developing countries for purposes of governance, decision-making, and monitoring of international programs, especially those related to development (Sanga et al. 2011, 303).

Even in cases where national statistical systems are able to generate a sufficient spectrum of needed data, events like the now famous 2010 Ghanaian GDP revision have cast a light of uncertainty over these figures, at least in sub-Saharan African nations (Jerven 2013). Studies have found that countries with the poorest quality statistics also have the fewest statistics (Williams 2011, 492). Accordingly, efforts to enhance national statistical capacity address both the scope and quality of statistical system operations. The evaluation of statistical capacity in developing countries has seen considerable growth over the last decade; there is emerging evidence that improvements in evaluation are leading to improvements in the systems (Willoughby 2008).

The international community has proven to be a lasting and strong supporter of efforts to evaluate and improve statistical capacity. This support emerged initially through setting standards, monitoring statistical operations and planning capacity building efforts. Soon, a number of international programs emerged to support further planning and implementation efforts. Examples include the Addis Ababa Plan of Action for statistical development (or AAPA), the World Bank's STATCAP program (Sanga et al. 2011) and the joint UN, OECD, IMF, World Bank and EC Partnership in Statistics for Development in the 21$^{st}$ Century or PARIS 21. There is at least some evidence that international programs are effective. For example, Alexander and his colleagues (2008) argue that increases in statistical efforts are associated with pursuit of the United Nations Millennium Development Goals.

It is true that there is considerable variation in the quality of national statistical systems. But what are the roots of these variations? More importantly, what are the factors that lead to reductions in statistical capacity? The World Bank has identified the following factors as threats to national statistical capacity: budget cuts, overdependence on donor financing, lack of training for statistical personnel, inadequate feedback from users of statistical information, and reluctance among government bureaucracies to embrace transparency that come hand-in-hand with statistical capacity (World Bank 2002). Experts in the field of economic development also suspect that poor quality and national statistical systems are at least partially rooted in the political sensitivity of statistics they are called upon to generate (Devarajan 2013).

Many of the problems facing national statistical systems are rooted in the structure of governance organizations and public agencies. Thus, problems of national statistical capacity are in some sense problems of public policy and administration. For example, decreases in national statistical capacity, especially in developing countries, are thought to deplete financial resources for government agencies. Resource constraints create what the World Bank has called a "vicious cycle, in which inadequate resources restrain output and undermine the quality of statistics, while the poor quality of statistics leads to lower demand and hence fewer resources" (World Bank, 2002, para 11). Complicating the effects of the so-called "vicious cycle" are the remnants of tumultuous economic times that spur difficult-to-monitor informal and black market economies (Jerven 2013; Gunter 2013).

While statistical capacity has been examined from a host of theoretical and practical perspectives, one aspect still unexamined is the role of technology in improving or threatening a nation's statistical capacity. Given the accumulated evidence relative to the effects of technology in organizations and markets, the hypothesis here is that increases in technological attainment are associated with increases in national statistical capacity. But what, specifically, are the causal mechanisms?

National statistical capacity is as much about the sharing of information as it is about the collection and validation of information. Technology facilitates information sharing by, among other mechanisms, institutionalizing information and exchange protocols (Yang and Maxwell 2011, Barua et al 2007). General purpose technologies (GPTs) can increase knowledge capital and help organizations innovate, but they may provide more benefit to some sectors than others, just as some countries may be more likely to adopt them (Helpman 1998, 2). This observation has considerable relevance to national statistical systems. One expert (Deaton 2001) notes the relatively simple threats to national statistical systems associated with measurement error in fieldwork among statistical workers. For a specific example, poorly coordinated field workers could incorrectly estimate poverty levels by missing houses (Deaton 2001, 134) or by double counting houses. Through supporting innovation, learning and knowledge sharing, GPTs can play a role in managing and enhancing the quality of field observation and verification. This view is supported by studies dating back decades that show increases in productivity associated with technology adoption in government organizations (Danziger 1979).

Technology may also improve the accuracy of sampling procedures in the field or facilitating communication. Other scholars provide insight to support indirect benefits of technologies. For example, it has been observed that difficult-to-monitor enterprises such as informal or unrecorded markets threaten the capacities of national statistical systems (Jerven 2013). It may be the case that access to general purpose technologies suppresses the growth of such difficult-to-monitor enterprises or increase the prospects of observing them. The concentration of informal markets, at least in relative terms, is greater in developing than developed countries. Developing countries often face rapid rates of social and economic change that stress governance mechanisms including national statistical systems. The empirical evidence in organizational studies shows that technologies increase organizations' capacity to adapt to change (Garcia-Morales et al. 2008). This may be in part to the increases in organizational learning that are known to be associated with technology (Bolivar-Ramoz et al. 2012).

This paper's focus on the level of technological attainment at the national level is particularly important because in the case of government organizations, there is evidence that the ability of an organization to benefit from technologies is at least in part dependent upon the technological sophistication of the regions in which the organizations are situated (Norris and Kraemer 1996).

**Model Specification**

To examine the hypothesis, this paper employs a cross-sectional time series dataset for the years 2004 to 2006 from a range of datasets on national statistical capacity and technological, economic, political, and regulatory characteristics. The dependent variable is the national statistical capacity of a country measured using the World Bank's Statistical Capacity Index or SCI. SCI is a multidimensional measure that takes into account the statistical methodology, source data, periodicity, and timeliness of a nation's statistical system. Scores range from a low of 0 to a high of 100 on three measures, with SCI being the mean of all three. The World Bank collects SCI annually for most developing nations. For each dimension, countries are scored according to specific criteria such that a score of 100 reflects a county that meets all criteria. Scores are made using data from a variety of sources including the World Bank, IMF, UN, UNESCO and WHO.[i]

Statistical methodology refers to a country's adherence to international methodological standards. Binary yes (1) or no (0) scores are given to countries based on adherence to ten specific methodological standards. Each score is given a weight of 10 for a maximum score of 100. A country with a score of 100 will have a national accounts base year (used for measuring

GDP) and a consumer price index (CPI) base year within the past 10 years (or have annual chain linking). It will also use the Balance of Payments Manual, have an actual or preliminary external debt reporting mechanism, it will have subscribed to the IMF's Special Data Dissemination Standard, and will have consolidated central government accounts. Finally, it will have reported key national measures (industrial production, import/export prices and vaccines) to a number of relevant international tracking organizations including the IMF, UNESCO and WHO.

Source data measures a country's frequency of obtaining statistical data such as population and poverty rate. Again, the highest score possible in this dimension is 100. Countries are measured according to five criteria, three of which are scored on a yes (1) or no (0) binary basis and two include scoring regimes that allow for half-points. Scores are assigned weights of 20. Countries that score 100 will have population and agricultural censuses every ten years or less, will conduct poverty and public health surveys every three years or less (for these two indictors half-point scores are awarded for surveys conducted every five years or less), and will have a complete vital registration system.

Periodicity and timeliness measures the availability of socioeconomic indicators, including those associated with the Millennium Development Goals. This component also rates the ease of access to key statistics. Scoring here is conducted according to 10 key indicators. Like the previous two dimensions, countries that meet the full criteria of an indicator are awarded a full point for a maximum of 10 points. Each point is assigned a weight of 10 for a maximum score of 100. Three indicators—periodicity of indicators for immunizations, HIV/AIDS, and child mortality—are scored on a binary yes (1) or no (0) basis. All other measures allow for partial point scoring in either halves or thirds. These other indicators include periodicity of indicators related to income poverty, child nutrition, maternal health, child education, access to water and GDP growth.

The differences between source data and periodicity and timeliness dimensions are important. The source data dimension measures a country's adherence to international standards relative to periodicity and also reflects whether administrative systems exist for the collection and estimation of key measures. The periodicity and timeliness dimension reflects a country's capacity to transform key socioeconomic source data into high quality useable data in a timely manner.

As noted above, the focus on technology in economics has led to increased study of technological variation across countries (e.g. Archibugi and Coco 2005). Most approaches center on assessing cross-national differences in people's and firm's access to GPTs and higher education. Technological capability is measured using the ArCo Technology Index (Archibugi and Coco 2004), a unified index that measures technological capabilities in developed and developing countries. The ArCo Index follows in the footsteps of earlier measures such as the United Nations Human Development Program's Technology Achievement Index and the United Nations Industrial Development Organization's Industrial Performance Scoreboard. ArCo improves on these measures by expanding the breadth of nations covered and emphasizing data that varies over time. The ArCo index combines eight measures (patents; scientific articles; Internet penetration; telephone penetration; electricity consumption; tertiary science and engineering enrollment; mean years of schooling; and, literacy rates). High index scores demonstrate advanced technological capabilities or attainment in the nation.

Generally, the index attempts to capture countries ability to create and diffuse technology; rather than seek to measure worldwide technology development, it measures countries level of participation in the creation and use of technology. The index captures a wide range of technological achievements and consolidates them into a single, comparable metric weighted towards the role of information.

Also included is an additional measure that assesses the impact of a country's regulatory environments on quality of the national statistical system. More wide-ranging than the capability of the government in domestic regulation, our measure addresses the World Bank's Governance Indicators estimate of each country's "regulatory quality."[ii] This measures the degree to which regulation is generally perceived by expert respondents relative to the market (e.g., in price controls, bank supervision, foreign trade, or business), and helps account for comparison problems across levels of development within our population of developing nations (e.g., Bertelli and Whitford 2009; Chinn and Fairlie 2006). Greater levels of regulatory quality are associated with stronger perceptions that the state engages in quality regulation of the market. It is recognized that national statistical systems vary in the extent to which they engage in monitoring and regulating market interactions. Accordingly, this paper predicts that increased perception of quality regulation is correlated with increased national statistical capacity.

Finally, the model considers whether statistical capacity varies based on the political environment. The democratization measure comes from the Polity IV dataset and considers institutional structures that focus on governmental authority patterns (Marshall, et al., 2002). Despite some criticism of this measure, it is chosen for its level of comparability across a range of countries. This measure as rescaled ranges from 0 to 20, with low values describing less democratization. A range of state governmental structures create a variety of motivators for political behavior. We hypothesize that democratization leads to greater statistical capacity. Indicators of ballot control and executive and legislative representation come from the Database of Political Institutions (Beck, et al. 2001). Measures of proportional representation (PR), open-list representation, and role of the executive are all dichotomous. Systems with fully independent presidents are classified as strong presidential systems in order to distinguish those from weak presidential and parliamentary systems. We expect that PR and open-list PR systems are more democratic; those countries will have enhanced statistical capacity. Likewise, we expect that PR systems will also have enhanced capacity, which is expressed through a negative sign for our two presidency variables. The last dichotomous variable is enrollment in the European Union. Table 1 shows the descriptive statistics for the data while Table 2 provides a list of countries included in the analyses.

**Table 1: Descriptive Statistics**

| Variable | 2004 | | 2005 | | 2006 | | Overall | |
|---|---|---|---|---|---|---|---|---|
| | Mean | SD | Mean | SD | Mean | SD | Mean | SD |
| Statistical Capacity Index (SCI) | 67.43 | 12.11 | 67.90 | 16.20 | 68.08 | 15.97 | 67.82 | 16.36 |
| Tech. Attainment | 2.60 | 1.2 | 2.59 | 1.20 | 2.59 | 1.20 | 2.59 | 1.2 |
| Regulatory Quality | -0.40 | 0.74 | -0.37 | 0.75 | -0.35 | 0.75 | -0.37 | 0.74 |
| Democratization | 12.95 | 6.21 | 13.11 | 6.14 | 13.17 | 6.15 | 13.08 | 6.15 |
| Strong President | 0.68 | 0.46 | 0.63 | 0.48 | 0.63 | 0.48 | 0.65 | 0.48 |
| Weak President | .120 | .326 | .127 | .334 | .117 | .323 | 0.12 | 0.33 |
| Proportional Representation | .597 | .493 | .567 | .497 | .567 | .497 | 0.58 | 0.49 |
| Open List | .403 | .492 | .394 | .490 | .403 | .492 | 0.4 | 0.49 |
| EU Membership | .054 | .228 | .048 | .215 | .048 | .214 | 0.05 | 0.22 |

**Table 2: Included Countries (n=94)**

| | | | |
|---|---|---|---|
| Albania | Egypt | Lithuania | Romania |
| Algeria | El Salvador | Macedonia | Russia |
| Argentina | Estonia | Madagascar | Rwanda |
| Armenia | Ethiopia | Malawi | Senegal |
| Azerbaijan | Fiji | Malaysia | Sierra Leone |
| Bangladesh | Gabon | Mali | Slovakia |
| Belarus | Georgia | Mauritania | South Africa |
| Benin | Ghana | Mauritius | Sri Lanka |
| Bolivia | Guatemala | Mexico | Sudan |
| Botswana | Guinea | Moldova | Syria |
| Brazil | Guyana | Mongolia | Tanzania |
| Bulgaria | Honduras | Morocco | Thailand |
| Burkina Faso | Hungary | Mozambique | Togo |
| Cambodia | India | Namibia | Trinidad And Tobago |
| Cameroon | Indonesia | Nicaragua | Tunisia |
| Central African Republic | Ivory Coast | Niger | Turkey |
| Chad | Jamaica | Nigeria | Turkmenistan |
| Chile | Jordan | Pakistan | Uganda |
| Colombia | Kazakhstan | Panama | Ukraine |
| Congo | Kenya | Papua New Guinea | Uruguay |
| Costa Rica | Kyrgyzstan | Paraguay | Venezuela |
| Croatia | Latvia | Peru | Yemen |
| Dominican Republic | Lebanon | Philippines | |
| Ecuador | Liberia | Poland | |

Estimation Results

The model evaluates the impact of technological capability on the national statistical capacity of 94 countries. To decrease the bias associated with time-dependent unobserved variables, the generalized linear model employs panel data from these 94 countries measured from 2004 to 2006 using the Generalized Estimating Equations (GEE) method, while accounting for typical traits of panel data including unobservable heterogeneity and serial correlation (Liang and Zeger 1986; Zeger and Liang 1986).[iii] This model is appropriate for cross-sectionally dominant data sets, yielding parameter estimates that are uncontaminated by heteroskedasticity and serial autocorrelation of errors (Zorn 2001). The underlying panel effects of repeated measures of the national statistical capacity may complicate estimation of the common coefficients. Variation of scale between units and variance within each panel make heteroskedasticity very likely. Accordingly, the model uses Huber-White standard errors to form a robust estimate of that variance. Finally, the model includes an AR(1) term to account for possible serial correlation. It is noted in advance that the sign and significance (and, to a degree, magnitude) of the effects we report for GPTs do not vary with specification; the results are robust to variations in the GEE correlation matrix, and even the use of alternative estimators such as tobit.

The dependent variable, national statistical capacity, is bounded [0,100]. Accordingly, the dependent variable is first calculated as a proportion ($\pi$), and then calculated by a logit transformation to create a new dependent variable (y = $\ln(\pi/(1-\pi))$). This allows estimation of the GEE model given the original bounding of the dependent variable.

Table 3 shows the GEE regression results for the model. The $\chi^2(7)$ statistic, 129.21, indicates that the model fits the data well. All models estimated include Huber-White robust standard errors.

## Table 3: GEE Estimates

| Variable | Estimate | Semi-Robust SE | |
|---|---|---|---|
| Tech. attainment | 0.3114 | 0.0536 | *** |
| Regulatory Quality | 0.2427 | 0.0901 | ** |
| Democratization | 0.0036 | 0.0076 | |
| Strong President | -0.1496 | 0.1296 | |
| Weak President | -0.2223 | 0.2168 | |
| Proportional Representation | 0.2187 | 0.1000 | * |
| Open List | 0.1024 | 0.1290 | |
| EU Membership | 0.0408 | 0.2655 | |
| Constant | 0.1585 | 0.1916 | |
| N | | 272 | |
| Wald $\chi^2(7)$ | | 129.21 | *** |
| Correlation Matrix | | AR(1) | |
| Scale Parameter | | 0.2990 | |

\*     $p < 0.10$ (two-tailed test)
\*\*    $p < 0.05$ (two-tailed test)
\*\*\*   $p < 0.01$ (two-tailed test)

**Figure 1: Estimated Effects of Technological Attainment on National Statistical Capacity**

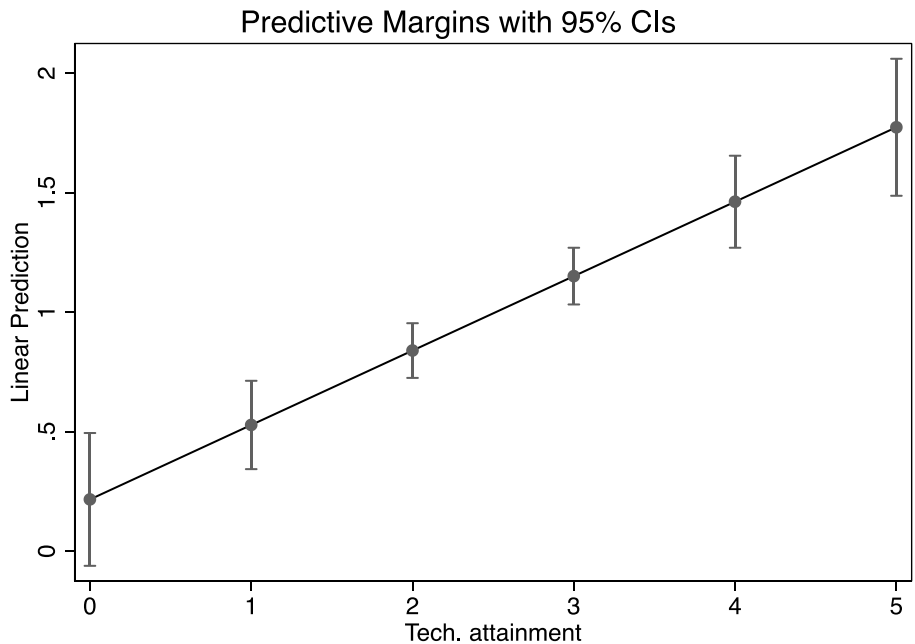

Table 3 presents the results. A review of the literature on growth theory and GPTs indicates that general technologies assist organizations in development of more specific technologies that lead to performance improvements; thus, greater use of general technologies provides an overall societal benefit. Together, these specific and general benefits support economic growth. The view taken here is that greater societal penetration of these GPTs allow nations to better monitor and measure themselves. This expectation (H1) is supported directly by the model. As Table 3 shows, the effect of a one-standard deviation shift in the ArCo Index is just under a one-standard deviation shift in the dependent variable. Figure 1 shows the statistically significant increase in national statistical capacity (including 95% confidence intervals) associated with increases in technological attainment.

However, technology is only one source of national statistical capacity. A higher incidence of national statistical capacity is observed with government systems that rely on proportional representation in their political systems. Findings also indicate a strong positive effect of regulatory quality on national statistical capacity. Figures 2 and 3 show the statistically significant increases in national statistical capacity (including 95% confidence intervals) associated with increases in regulatory quality and proportional representation, respectively. There is no evidence of direct effects from democratization, European Union membership, or presidential systems, which undermine our arguments about political attributes. None of these are associated with greater statistical capacity. Findings do indicate an effect for PR systems, with having a PR system causing about a one-third of standard deviation improvement in the dependent variable.

**Figure 2: Estimated Effects of Regulatory Quality on National Statistical Capacity**

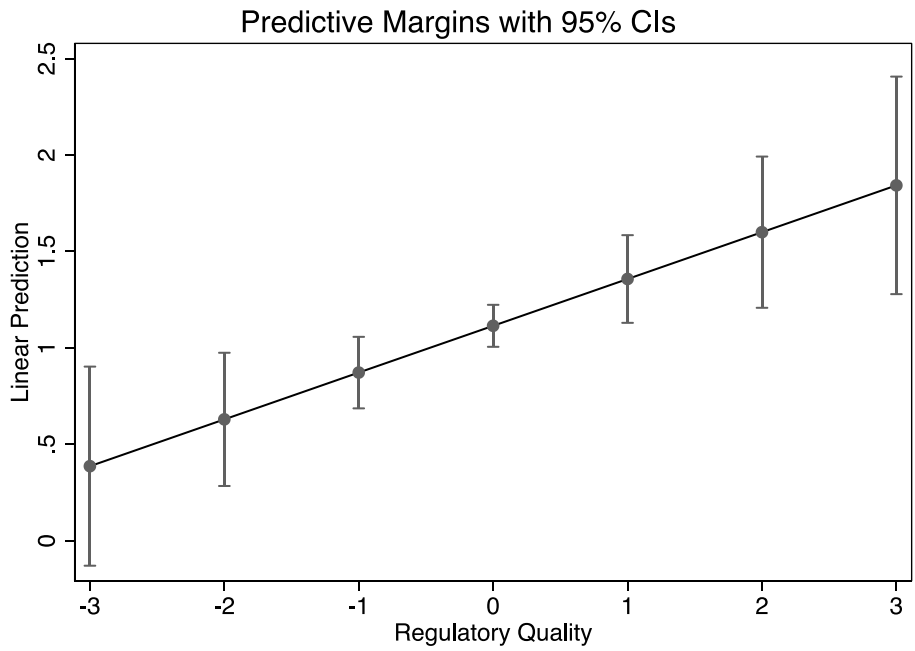

**Figure 3: Estimated Effects of Proportional Representation on National Statistical Capacity**

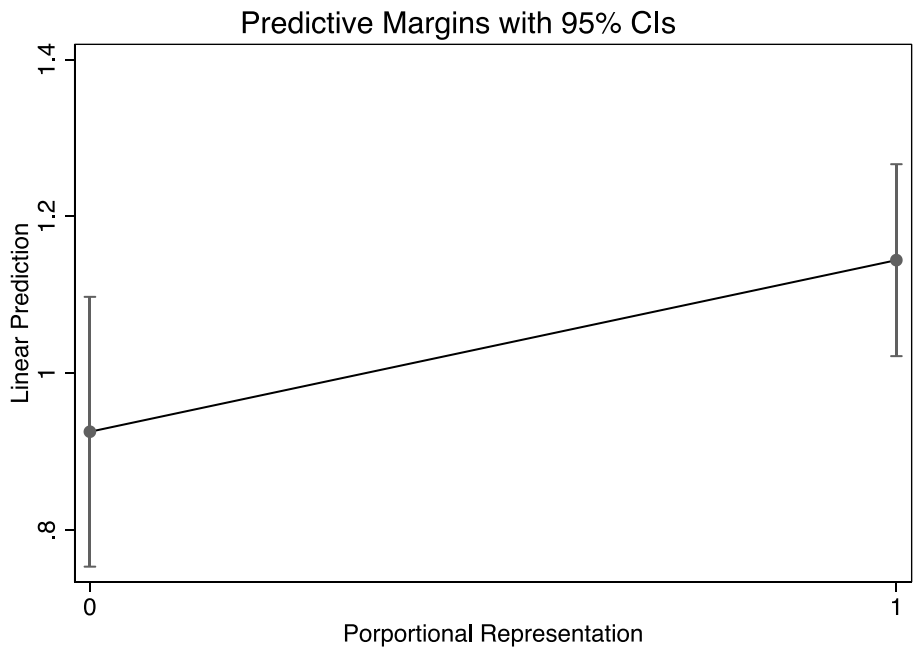

The results shown here demonstrate a significant impact of technological attainment on national statistical capacity. Since technological capability is generally not controlled by the government or policy designers, what can they do to improve statistical capacity? Is it fruitless for them to attempt progress toward effective national statistical systems? One view is that general technologies serve multiple purposes, including economic growth, so organizations and governments adapt to new technologies for financial reasons. Inferences relative to traditional causes such as political systems can be reinterpreted upon accounting for varying levels of technological attainment across nations; technology is more important than factors like European Union membership or democratization.

## Discussion

Effective systems of government increasingly rely on knowledge and evidence. National statistical systems play an important role in systematically collecting, validating and disseminating important knowledge and evidence at the heart of many important government enterprises. It has been observed that the problems facing national statistic systems are those of both knowledge and governance (Jerven 2013). Governments, especially those in developing countries, rightly bent on improving performance, are likely to turn increasingly to their national statistical systems to collect, validate and distribute the information they need for governance and decision making. However, as manifested in the now famous 2010 Ghanaian GDP revision, there is reason to be skeptical of national statistics, especially those produced by developing countries (Jerven 2013, Devarajan 2013). Further demand for high quality national statistics is fueled by a proliferation of international economic development and poverty reduction programs. In as much as statistics are the so-called "starting point" in the war on poverty (World Bank 2002), there will be increased emphasis on improving the capacity of national statistical systems. Within this context, questions emerge relative to the best channels for fostering meaningful improvements.

The response from the international development community to the "statistical tragedies" (Devarajan 2013) facing developing countries has been to double down on programs to evaluate (e.g., Willoughby 2008), plan for, and strategically design improvements in national statistical systems. Programs such as the World Bank's STATCAP project or PARIS21 seem to flourish. While these programs appear to be effective, insights from scholars remind the development community that national statistical systems exist in diverse economic, social and, political environments where one size fits all approaches are hard pressed to succeed. Little attention has been placed on the domain of possible factors contributing to the improvement of national statistical systems but remaining largely outside the control of government. Technological attainment is one such factor.

The aim of this study is to add to what is known about determinants of national statistical capacity. The focus is on technological attainment as a foundation for national statistical capacity. Economists have recognized cross-national differences in technological attainment as a primary cause for variation in economic growth (e.g., Helpman 1998; Islam 2003; Castellacci 2007). Technology also contributes to the development of governance practices and the regulatory state (Whitford and Tucker 2009). Accordingly, the evidence presented here is that nations with more widespread use of "general purpose technologies" (GPTs) have greater national statistical capacities. The effects of GPTs on organizations and economic development are well documented. As innovations, GPTs may find uses in numerous economic sectors and drive change in the operations of many organizations (Helpman 1998, 3); they serve as "enabling technologies, allowing organizations to exploit market complementarities and select new modes of action (Bresnahan and Trajtenberg 1995).

Findings indicate that technological attainment is a strong predictor of national statistical capacity. These findings complement conventional thinking on the benefits of technology to economies and organizations. Moreover, the evidence on the benefits of national statistical capacity to systems of governance provide a compelling case for, in the very least, massive international investment in bolstering the effectiveness of national statistical systems. These findings relative to the benefits of technology may provide a powerful compliment to such efforts.

**Notes**

[i] For more information on SCI see World Bank (2014).

[ii] The index considers multiple independent surveys and results based on a detailed measurement model that is described by Kaufmann and colleagues (2008).

[iii] OLS estimations of models using pooled cross-sectional time-series data often violate assumptions of homoscedasticity and error term correlation (Kmenta 1986). While OLS estimated coefficients are unbiased in the presence of autocorrelation, these estimates are not efficient, and OLS coefficient variability threatens assessments of statistical significance.

Devarajan, S. 2013. "Africa's statistical tragedy." *Review of Income and Wealth* 59(S1): S9–S15.

Dosi, G. 1988. "Sources, procedures, and microeconomic effects of innovation." *Journal of Economic Literature* 1120–1171.

DuBose, J., Frost, J. D., Chamaeau, J. A., and Vanegas, J. A. 1995. "Sustainable development and technology." *The environmentally educated engineer* 73-86.

Fagerberg, J. 1994. "Technology and international differences in growth rates." *Journal of Economic Literature* 32(3): 1147–1175.

García-Morales, V. J., Lloréns-Montes, F. J., and Verdú-Jover, A. J. 2008. "The effects of transformational leadership on organizational performance through knowledge and innovation." *British Journal of Management* 19(4): 299–319.

Gössling, S., Borgström Hansson, C., Hörstmeier, O., and Saggel, S. 2002. "Ecological footprint analysis as a tool to assess tourism sustainability." *Ecological Economics* 43(2–3): 199–211.

Grossman, G. M. and Elhanan, H. 1994. "Endogenous innovation in the theory of growth." *Journal of Economic Perspectives* 8(1): 23-44.

Gunter, S. (2013). "State Earned Income Tax Credits and Participation in Regular and Informal Work." *National Tax Journal* 66(1): 33–62.

Helpman, E. (Ed.). 1998. *General purpose technologies and economic growth*. Cambridge, MA: The MIT Press.

Islam, N. 2003. "What have we learnt from the convergence debate?" *Journal of Economic Surveys* 17(3): 309–362.
28